# Shot-noise-driven macroscopic vibrations and displacement transduction in quantum tunnel junctions

Prasanta Kumbhakar[1], Anusha Shanmugam[1], Akhileshwar Mishra[1], Ravi Pant[1], J L Reno[2], S Addamane[2], and Madhu Thalakulam[1*]

[1] Indian Institute of Science Education & Research Thiruvananthapuram, Kerala 695551, India

[2] Center for Integrated Nanotechnologies, Sandia National Laboratory, Albuquerque, NM, USA

## Abstract

Inherent randomness and the resulting stochastic behavior of fundamental particles manifested as quantum noise put a lower bound on measurement imprecision in the quantum measurement process. In addition, the quantum noise imparts decoherence and dephasing to the system being measured, referred to as the measurement back-action. While the microscopic effects of back-action have been observed, macroscopic evidence is a rarity. Here we report a macroscopic display of the back-action of an ultra-sensitive quantum point contact (QPC) electrical amplifier whose transport is defined by the quantum tunneling of electrons. The QPC amplifier, realized on GaAs/AlGaAs heterostructures, coupled to a planar superconducting resonator, operates at a frequency of $\sim 2.155$ GHz in the shot-noise-limited regime. The shot-noise excitation of the mechanical modes and the resulting piezoelectric polarization enhancing the shot-noise at the mode frequencies form a positive feedback loop between the electrical and mechanical degrees of freedom. While the excitation of the vibrational modes is a display of the macroscopic effects of measurement back-action, the amplitudes of the noise peaks allow us to calibrate the displacement sensitivity of the QPC-resonator systems, which is in the $\sim 35$ fm/$\sqrt{\text{Hz}}$ range, making it an excellent sensor for ultra-sensitive and fast strain or displacement detection.

[*] madhu@iisertvm.ac.in

Wave-particle duality and quantum uncertainty are the underlying doctrines governing the behavior of quantum systems. The resulting discreteness and uncertainty assert a statistical nature, manifested as shot-noise in the field of sensitive quantum measurements, dictating the sensitivity of the measurement [1]. Besides limiting accuracy, the noise has detrimental effects on the state of the system being measured, generally referred to as the measurement back-action [1–3]. While both quantum and classical processes can contribute to the noise and back-action, the classical sources yield only the equilibrium energy, the temperature of the system [4]. In contrast, shot-noise can also reveal valuable information not easily accessible otherwise, such as the governing statistics, interparticle correlations, and bunching-antibunching mechanisms present in the system [5–8]. Despite the fact that shot-noise was initially seen on electrons moving in vacuum [9], the dominant contributions from other noises and interactions present make the observation of electronic shot-noise in solids a nontrivial problem. The origin being randomness, quantum devices exhibiting tunneling-dominated transport are the best candidates to observe shot-noise. Development in material-growth, sophisticated fabrication, and measurement techniques enabled the advent of mesoscopic solid-state circuits exhibiting shot-noise governed transport; quantum point contacts (QPC), atomic point contacts, and semiconductor quantum dots [10–16] are the important ones. Remarkably, these devices also make ultra-sensitive detectors [17–19]. Among them, QPCs are one of the most sought-after for a variety of technologies, such as, qubit readout [20], charge counting in electrical metrology circuits [21,22], and displacement sensing in electromechanical systems [23,24].

Back-action imparted on the system being measured by these devices is a major cause of concern, especially in qubit readout and quantum error correction schemes. The broadband tunneling shot-noise imparted during measurement has been shown to affect fidelity [25–27], induce dephasing, and drive inelastic charge transitions on coupled devices [28–31]. In hybrid systems such as opto- and electro-mechanical devices, owing to the positive feedback and interplay between coupled degrees of freedom, the back-action can lead to phenomena, both fundamentally and technologically non-trivial. In micro-electromechanical systems, the tunneling shot-noise driving mechanical vibrations and the vibrational modes altering the shot-noise spectra are observed [2,32–34]. In optomechanical systems, not just at the microscopic level, the photon shot-noise perturbing macroscopic objects, driving mechanical vibrations, and affecting the measurement process has been reported [3,35]. Such macroscopic effects of electron shot-noise are seldom observed [34].

In this work, we report the macroscopic manifestation of the back-action of electron tunneling in an ultra-sensitive electrical amplifier consisting of a QPC embedded in a superconducting microwave cavity, operated at 10 mK. From the power spectral density of the noise and signal, we confirm that the amplifier operates within the shot-noise limit. Being realized on a piezoelectric platform, the positive feedback between the electrical and mechanical degrees of freedom facilitates (i) the excitation of vibrational modes driven by shot-noise and (ii) vibrational modes inducing polarization charges, enhancing tunneling across the QPC, and modulating the shot-noise spectra. From the noise power spectra, we estimate a displacement sensitivity of $\sim 35 \text{ fm}/\sqrt{\text{Hz}}$. The results discussed in this work, besides improving our understanding of quantum back-action, also have an impact on the realization of quantum electrical amplifiers by shot-noise engineering. The ultra-high displacement sensitivity also makes this system a promising candidate for applications such as gravitational wave detection.

The device consists of a GaAs/AlGaAs QPC coupled to an Aluminium superconducting coplanar waveguide (CPW) resonator (see Supplementary Information S-1 for details on device fabrication and experimental setup) [36]. The QPC and the CPW resonators are fabricated using standard microfabrication processes. An optical image of a resonator similar to the one used in this experiment is provided in Supplementary Information S-2. The high-frequency response of the system is measured using the RF reflectometry technique [36,37]. Fig. 1(a) shows a schematic diagram of the measurement configuration. Fig. 1(b) shows the transconductance of the QPC exhibiting a sharp pinch-off of the constriction and subsequent shut-down of the channel current for gate voltages less than $V_g \sim -0.8$ V. The inset to Fig. 1(b) shows an SEM image of the QPC gates.

The resonator described in Fig. 1(a) exhibits multiple resonating modes; $\lambda$, $3\lambda/4$ and $5\lambda/2$ modes at frequencies of 3.03 GHz, 2.17 GHz, and 2.276 GHz, respectively, are shown in Supplementary Information S-3 and S-4. Though the reflectivity of the resonator at all these resonances is a function of the QPC transconductance [36], the $3\lambda/4$ mode at 2.17 GHz, exhibits a sharper response than the other two modes. Fig. 1(c) shows $S_{11}$ for various QPC transconductances, represented by solid squares in Fig. 1(b). The large variation in the resonator reflected power against the QPC conductance is exploited to realize an ultrasensitive electrical amplifier. The sensitivity of the QPC is determined using the formula $\delta G = \frac{1}{2}\frac{\Delta G}{\sqrt{BW}10^{SNR/20}}$, where $\Delta G$ is the small signal conductance change, $BW$ is the resolution bandwidth, and $SNR$ is the signal-to-noise ratio of the sidebands generated in the reflected power spectrum. We find that the SNR peaks at ~ 2.155 GHz. The data described in the remaining portion of the manuscript are acquired at this carrier-wave of frequency $f_c$ = 2.155 GHz unless specified otherwise. We obtain a best conductance sensitivity of ~ $3.7 \times 10^{-4}$ ($e^2/h$)/$\sqrt{Hz}$ [see Supplementary Information S-5 for details].

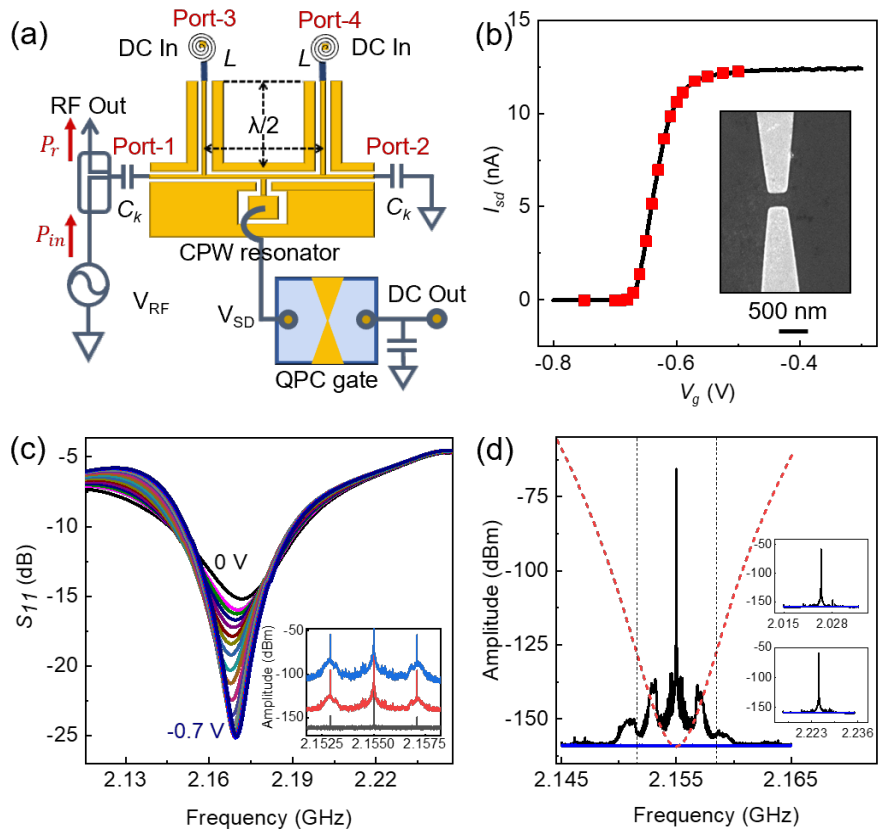

FIG. 1. (a) Schematic diagram of the device configuration. (b) QPC transconductance measured at 10 mK. (Inset: SEM image of a QPC device similar to the measured one.) (c) $S_{11}$ of the resonator-QPC system for various QPC transconductance values, shown in red-squares in panel-(b). Inset: Small signal response with a 2.1 MHz signal applied onto the QPC gate $f_c$ = 2.155 GHz for $P_{rf}$= -84 dBm (bottom), -51 dBm (middle) and -37 dBm (top). (d) Reflected power spectrum in a span of 20 MHz around the carrier wave (2.155GHz) for $P_{rf}$= -44 dBm. Insets: Reflected power spectra for $f_c$ = 2.025 GHz (top) and 2.225 GHz (bottom). Blue-trace: noise floor of the cryogenic HEMT amplifier.

Inset to Fig. 1(c) shows the small signal response of the system for $P_{rf} = -84$ dBm (bottom), $-51$ dBm (middle) and $-37$ dBm (top), for a QPC gate excitation signal $f_m = 2.1$ MHz with the QPC is biased at $V_g \sim -0.67$ V. We find that as the $P_{rf}$ is increased, the noise floor of the spectra develops broad peak-like structures in addition to the applied excitation.

To probe this further, we inspect the reflected power spectrum in the vicinity of $f_c$ in a wider frequency span of 20 MHz [Fig. 1(d)]. When the device is operational, we notice the appearance of a set of spontaneously developed broad peaks at frequencies ~560 kHz, ~2.15 MHz, and ~4.3 MHz, in addition to an overall rise in the noise floor above that of the secondary amplifier (blue trace). The peaks subside substantially when we drive the system away from the resonance, as shown in the insets to Fig. 1(d) top ($f_c = 2.025$ GHz) and bottom ($f_c = 2.225$ GHz). Identical spectral responses are observed at other resonances [Supplementary Information S-6], suggesting these features are independent of the carrier frequency.

The $1/f$ noise and the thermal noise contributions are negligible for the operational frequency and temperature ranges discussed in this work. An important noise source is the QPC itself, which produces two kinds of noise: (i) charge noise coupled to the channel and (ii) shot-noise [4,38]. Spurious charge noise will behave like a modulation signal, and the power enclosed will be $\propto P_{rf}$. Fig. 2(a) shows the power spectra of the spontaneously developed peaks in the reflected signal while the QPC is biased at $V_g \sim -0.675$ V. The peaks show a growth as $P_{rf}$ is increased. Fig. 2(b) shows a plot of the integrated noise power $P_n$ versus $P_{rf}$, in a 1.3 MHz frequency span enclosing the ~2.15 MHz peaks (open-triangles) and 5 MHz span (open-squares) covering all the noise peaks on the left-half, excluding the carrier wave. These frequency ranges are marked by the double-headed arrows in Fig. 2(a). The power in a side-band due to a small gate excitation of 3 mV$_{RMS}$ at 10 kHz (open-circles) shown in Fig. 2(b) varies as $P_{rf}$, in accord with the theory of amplitude modulation. In contrast, $P_n$, for nearly three decades of $P_{rf}$, varies as $\sqrt{P_{rf}}$. This points to the fact that the generator of the noise spectra is inherent to the transport mechanism in the QPC, and the $\sqrt{P_{rf}}$ dependance points towards shot-noise.

The relevant form of shot-noise here is photon-assisted shot-noise (PASN). The electron-hole pairs generated by the RF photons undergo partitioning at the QPC, producing the shot-noise [14,34]. The RF voltage $V_{rf}$ across the QPC for the range of $P_{rf}$ is shown in Fig. 2(b) inset (refer Supplementary Information S-7). For $\sqrt{2eV_{rf}}/\hbar\omega_c \gg 1$ and for the temperature range discussed in this report, the power spectral density of PASN scales as $\sqrt{P_{rf}}$ and as $T_n(1 - T_n)$ [4,39]. $T_n$ is the transmission probability of $n^{th}$ mode at the constriction. Fig.

2(c) shows the measured $P_n$ in a band of 1.3 MHz enclosing the ~ 2.15 MHz peak as a function of the QPC conductance (right) for $P_{rf} = -54$ dBm; as guide to the eye, the QPC current is shown on the left axis. In Fig. 2(c), the panels on the right show representative noise spectra while the constriction is pinched off (region-I), transport dominated by tunneling (region-II), and when the constriction is fully open (region-III). From $P_n$ and the noise spectra, we infer that the noise is negligible when the channel is completely open and pinched off ($T_n = 0$ or 1), and maximum when the transport is dominated by tunneling, i.e., in the middle of the pinch-off region, where $T_n \approx 0.5$. Except for the frequency dependence, the fundamental behavior of the noise power spectra we observe agrees with the theory of PASN. The observed spectral features suggest that the tunneling process across the QPC has been enhanced by an additional AC component.

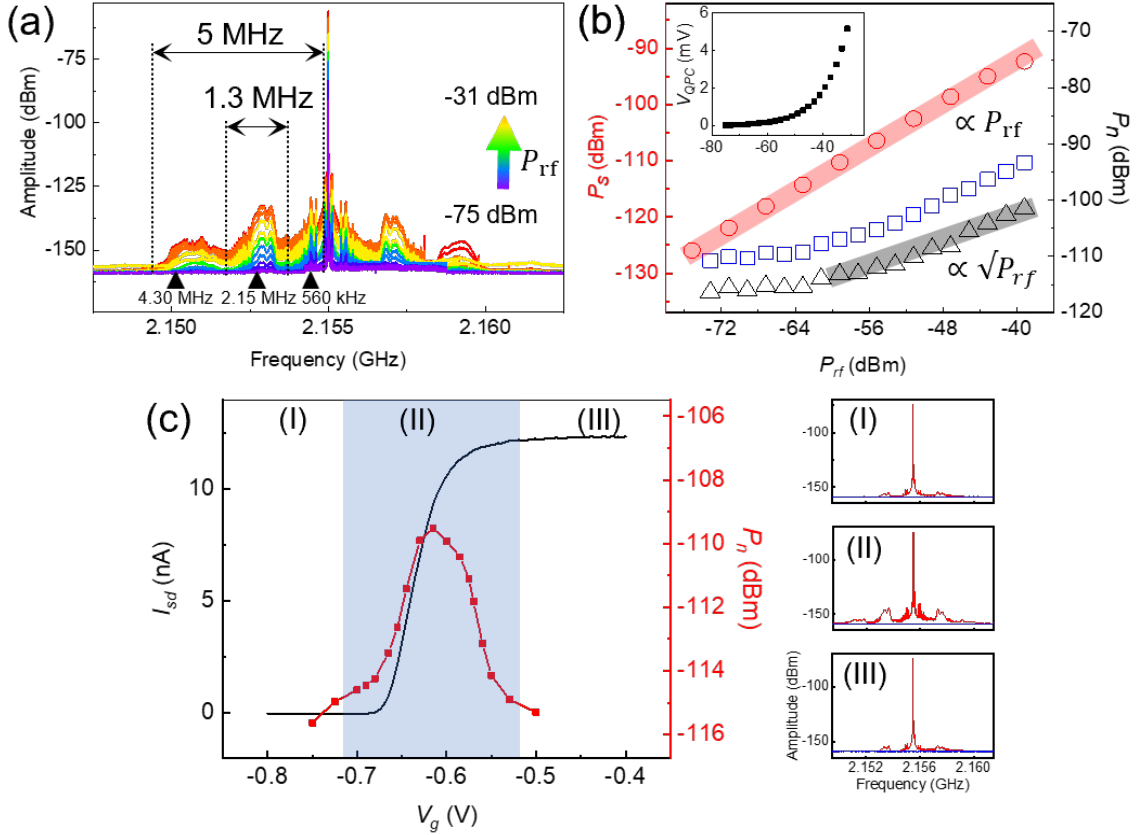

FIG. 2. (a) Reflected power spectrum of the resonator-QPC system for various $P_{rf}$ values at $V_g \sim -0.67$ V. (b) Signal power $P_s$ (red circles, left axis) vs. $P_{rf}$ for $f_c$=2.155 $GHz$, $f_m$=10 kHz and gate excitation ~ 3 mV$_{RMS}$. Right axis: Integrated excess noise power $P_n$ vs $P_{rf}$ for the frequency band 5 MHz (open squares in blue) and 1.3 MHz (open triangles in black). The grey-shaded region marks the range where $P_n$ scales as $P_{rf}^{1/2}$. (c) $P_n$ (right axis) vs. $V_g$; with $P_{rf} = -54$ dBm and $f_c$=2.155 GHz (left axis). The transconductance of the QPC is shown in black (left axis) as a guide to the eye. The shaded portion highlights the tunneling regime. (I), (II) & (III) Reflected power spectrum at three regimes of the transconductance trace.

Being a piezoelectric material, the electrical and mechanical degrees of freedom are coupled in our sample. When the tunnel rate $\Gamma \gg \nu_m$, the mechanical resonant frequency, the mechanical modes experience many piezoelectrically transduced random kicks during one oscillation period, causing a net energy transfer from the shot-noise to the mechanical mode, driving the piezoelectrically active vibrational modes [40–42]. On the other hand, the piezoelectrically active mechanical modes can cause a potential difference across the QPC, enhancing the tunneling current and thus the shot-noise at those modal frequencies. In the remaining sections, we show how the positive feedback loop formed by these two mechanisms generates peaks in the noise spectra [34].

The GaAs/AlGaAs crystal on which the QPC is of (100) orientation and the transport direction is along the (110) direction. Piezoelectrically active flexural and thickness-shear modes take part in the electromechanical interaction [43,44]. Consider the flexural mode $Ey-1$ as illustrated in Fig. 3(a) left panel. Around the constriction, this mode will cause a compression (outer edge) and an expansion (inner edge) along the z-direction, $\Delta z$. The resulting strain $S_4 = 2\Delta z/w$ will result in a polarization $P_1 = e_{14}S_4$ and bound charge formation $\Delta n$. $w$ is the width of the crystal. The bound charges will be screened by the 2DEG except in the depleted region under the gates and around the constriction. Across the constriction, the potential difference $\Delta\varepsilon$ due to $\Delta n$ modulates the tunneling current at frequencies prescribed by the mechanical modes. There is excellent agreement between the frequencies of the simulated piezoelectrically active modes and the observed spontaneously generated noise peaks, supporting our inference (see Supplementary Information S-8). Among the mechanical modes, even-numbered modes will not induce much strain at the location of the QPC, and as a result, those frequencies will not show up in the noise spectrum, $Ey-2$ mode for instance.

At the location of the ohmic contacts, polarization has a dipolar component in the z-direction [34], which helps to activate the mechanical modes, completing the feedback loop. For GaAs, the activation power for the piezoelectric vibration is low ~ µW [45]. A fluctuating charge $\Delta n$, can induce a force $\Delta F = \eta \Delta n$, where $\eta = \frac{\varepsilon_r - 1}{\varepsilon_r} \frac{\omega_m^2 ml}{c_{44} d_{14} w_c^2}$ is the piezoelectric force constant [34], $m$ is the mass of the crystal, $w_c^2$ defines the area of the ohmic contact and $\omega_m = 2\pi\nu_m$ is the resonance frequency of the vibrational mode. For our sample, $\eta = 7.29 \times 10^{14}$ N/C suggesting a small charge fluctuation can induce a substantial mechanical force on to the crystal. The crystal oscillator can easily be driven by the back-action force due to the shot-noise induced charge fluctuations. These modes induce an oscillatory polarization field and the polarization charge $\Delta n$. The combined effect of both halves of the feedback loop alters the white color of the shot-noise. We have made similar observations on another QPC device with slightly different spatial dimensions and operating frequency [not shown]. From the simulations, we find that the observed noise peaks show a very good correlation with the macroscopic dimension of the sample.

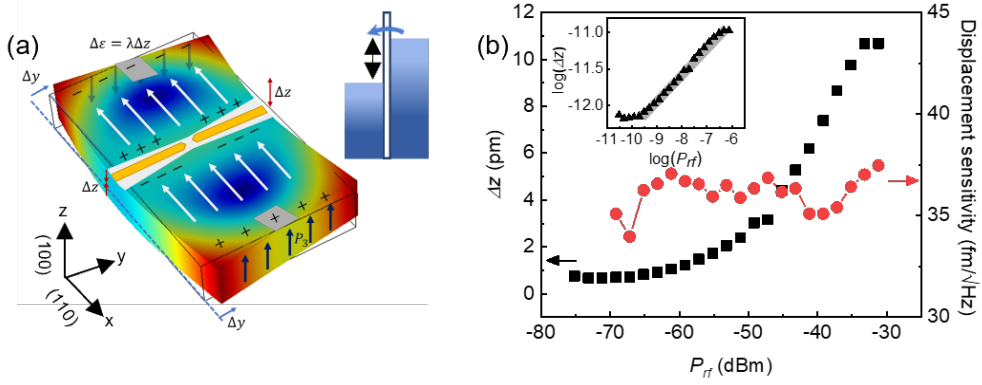

FIG. 3. (a) COMSOL simulation showing the bending of the GaAs crystal overlayed with a cartoon of the QPC and bound charge formation. $\Delta y$ and $\Delta z$ are the deformation at edges and middle of the chip respectively. Right: chemical potential difference $\Delta \varepsilon$ across QPC caused by the polarization charges. (c) Displacement $\Delta z$ as a function of $P_{rf}$ (left axis) and displacement sensitivity vs $P_{rf}$ (right axis). Inset: The log-log plot of $\Delta z$ vs $P_{rf}$. The shaded line represents a slope of 0.33

The noise spectra in the reflected power resulting from the mechanical modes can be directly used to sense displacement or strain. Considering a damped harmonic oscillator model for these modes, one can calculate the mean square voltage across the QPC due to the mechanical mode as $\langle V^2 \rangle = \frac{\pi v_m}{2Q_m} V_{n,max}^2$ where $v_m$ and $Q_m$ are the resonant frequency and quality factor of the mechanical mode [34]. The peak noise voltage $V_{n,max}$ and power $P_{n,max}$ are related by $V_{n,max}^2 = \frac{P_{n,max}}{G_{QPC} K\, RBW}$. $RBW$ is the resolution bandwidth, $G_{QPC}$ is the QPC conductance and $K$ the resonator-QPC power-coupling factor. For piezoelectrical coupling, $\Delta z = \Delta \varepsilon / \lambda = eV/\lambda = (e/\lambda) \sqrt{\frac{\pi v_m}{2Q_m} \frac{P_{n,max}}{G_{QPC} K\, RBW}}$. $\lambda \approx 1.5$ fN for our sample [see Supplementary Information S-9 for calculation details]. Since the frequency of $Ey - 1$ mode, ~560 kHz, is very close to the driving frequency and is influenced by the rise in the background, we conduct all our analysis using the $Ey - 3$ mode ~2.15 MHz. The noise spectra corresponding to different modes consists of multiple peaks, which we believe is due to the non-uniform sample geometry and the ohmic contacts. The peak heights and $Q_m$ are extracted from the spectral decomposition as shown in the Supplementary Information S-10.

Fig. 3(b) shows a plot of the displacement $\Delta z$, extracted from the power in the noise peak located at ~2.15 MHz as a function of $P_{rf}$. For shot-noise-generated vibrational modes, the displacement $\Delta z \propto \sqrt{P_{n,max}} \propto P_{rf}^{1/4}$. The inset shows a log-log plot of the same data with a slope of ~ 0.33 (indicated by the shaded line) agreeing satisfactorily with the predicted value of ¼. The red-trace in Fig. 3(b), right-axis, shows calculated displacement sensitivities, which are in the range of ~ 35 fm/$\sqrt{Hz}$, making this system a highly sensitive displacement or strain sensor.

Now we inspect the presence of thermally activated vibrational modes. Fig. 4(a) shows $P_n$ versus $P_{rf}$ acquired at various temperatures between 200 mK and 1.5 K. The traces in the lower-right inset to Fig. 4(a) show a magnified view of the noise spectra around ~2.15 MHz versus temperature. From Fig. 4(a), we find that for temperatures up to ~ 800 mK, the noise spectra show little variation with temperature. Similarly, the traces obtained for 1.2 K and 1.5 K are also nearly identical. When the temperature was raised from 800 mK to 1.2 K, a sudden downward shift in the overall noise power is observed, negating any contribution from the thermally excited modes. The nearly identical noise floor for the secondary amplifiers for all these temperatures, left-inset, Fig. 4 (a), also suggests little thermal-noise contribution to the measurement system. The filled blue circles show the behavior of the noise in the temperature range between 1 K and 1.1 K. The reduction in the shot-noise power as the system is taken from 800 mK through 1.2 K suggests a drop in the power transferred either from external RF circuit to the system or from the resonator to the QPC.

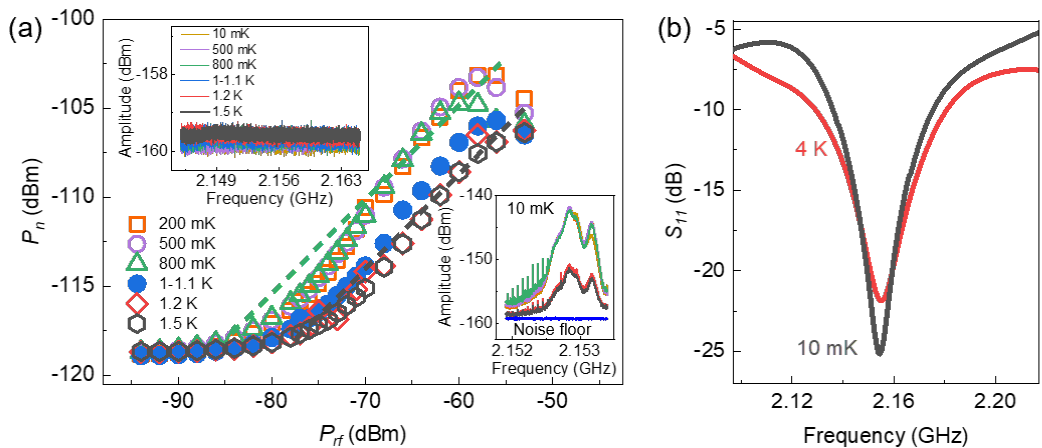

FIG. 4. (a) Integrated noise power $P_n$ in a span of 1.3 MHz around the 2.15 MHz peak (shown in the lower right inset) versus $P_{rf}$ for temperatures between 200 mK and 1.5 K. Upper-left inset: noise floor of the secondary amplifiers against temperature. Lower-right inset: 2.15 MHz peaks against temperature. (b) $S_{11}$ at 10 mK (black) and 4 K(red).

The $S_{11}$ traces shown in Fig. 4(b) suggest that the power transferred to the system is ~3 − 4 dB lower when the system was taken from 10 mK to 4 K mostly due to the CPW resonator, made of Aluminium, turning normal from the superconducting state at ~1.1K. The intermediate noise power exhibited by the trace corresponding to the transition regime (1 K - 1.1 K, Fig. 4(a) filled circles) corroborates with this picture. Given the measurement uncertainties, the behaviour of the traces shown in Fig. 4(a) also match quantitatively with the $S_{11}$ data shown in Fig. 4(b). The traces in Fig. 4(a) are shifted leftward by ~ 4 − 5 dB suggesting that one needs to increase $P_{rf}$ by that value to maintain the same noise power density as we take the system from 800 mK to 1.2 K. The power transferred from the resonator to the QPC is not expected to change significantly since the change in the resonator resistance compared to the resistance of the QPC is insignificant.

In this manuscript, the back-action of electron-tunneling events exciting macroscopic vibrations is demonstrated, in addition to a QPC-based shot-noise-limited electrical amplifier. The positive feedback and power transfer between the mechanical and electrical degrees of freedom modify the flat shot-noise spectra, enhancing tunneling at frequencies dictated by the mechanical modes. The ability to respond to minute mechanical displacements caused by the vibrational modes also makes this system an ultra-sensitive strain/displacement sensor operating in the quantum regime. The bunching of shot-noise onto the frequency bands dictated by the macroscopic modes, as demonstrated in this manuscript, lays down a recipe to reach the quantum limit in semiconducting charge/electrical amplifiers by engineering the device geometry. Shot-noise excitation of vibrational modes also has serious consequences in technologies where transport is dominated by tunneling, such as quantum dot qubits and quantum metrological circuits.

**Acknowledgements**

MT acknowledges the funding received from DST, Govt. of India under the grand ST/ICPS/QuST/Theme-4/2019 and MoE-STARS under the grand MoE-STARS/STARS1/363 and Prof. Yigal Meir for scientific discussions. PK acknowledge CSIR, Govt. of India for fellowship. This work was performed, in part, at the Centre for Integrated Nanotechnologies, an Office of Science User Facility operated for the U.S. Department of Energy (DOE) Office

# Supplementary Information

# Shot-noise-driven macroscopic vibrations and displacement transduction in quantum tunnel junctions

Prasanta Kumbhakar[1], Anusha Shanmugam[1], Akhileshwar Mishra[1], Ravi Pant[1], J L Reno[2], S Addamane[2], and Madhu Thalakulam[1*]

[1] Indian Institute of Science Education & Research Thiruvananthapuram, Kerala 695551, India

[2] Center for Integrated Nanotechnologies, Sandia National Laboratory, Albuquerque, NM, USA

## S-1: Device fabrication and Experimental setup

**Device fabrication**: The device consists of a QPC galvanically coupled to a superconducting CPW resonator. The QPC is formed on a GaAs/AlGaAs heterostructure hosting a two-dimensional electron gas (2DEG) $\sim 225\ nm$ below the crystal surface, with a carrier concentration of $\sim 2 \times 10^{11}\ cm^{-2}$ and a mobility of $\sim 2.13 \times 10^6\ cm^2 V^{-1} s^{-1}$. Ohmic contacts to the 2DEG are realised by Indium alloying technique while the surface gates defining the constriction is realised by electron-beam lithography followed by Cr/Au metallization. The CPW resonator is fabricated on a Sapphire substrate using photolithography followed by Aluminum deposition. The resonator has a lithographic length of $\sim 4.25$ cm defined by $\sim 10$ μm gaps in the center conductor. The DC and the low-frequency AC excitations are introduced onto the QPC source electrode via symmetrically placed, inductor-terminated feedlines of length $\lambda/2$ spliced to the center conductor at distances of $\lambda/4$ from the coupling capacitors on either end of the resonator. The source contact of the QPC is connected to the center of the resonator via bonding pads, while the drain is grounded to the ground plane of the resonator through a capacitor of 50 nF capacitance allowing grounding high-frequency signals to the ground plane of the resonator, while returning the DC and low-frequency current to the room-temperature current pre-amplifier. A home-built bias tee is used to apply high-frequency signals along with the DC gate bias to one of the QPC gate electrodes. The entire device is mounted in a sample box equipped with cryogenic high-frequency filters on the mixing chamber plate of a cryogen-free dilution refrigerator with a base temperature of 10 mK.

**Experimental setup**: DC and low-frequency AC measurements across the QPC are performed in the two-probe configuration. The current returned from the device is measured using a low-noise current pre-amplifier (SR570). RF measurements are performed in the radio-frequency reflectance mode. RF signals at the resonance frequency are sourced using RF signal generator.

* madhu@iisertvm.ac.in

The reflected signals from the CPW resonator input port are redirected using a directional coupler situated at the 100 mK temperature stage to a cryogenic high-frequency amplifier (Cosmic microwave) with a noise temperature of 7 K and a gain of ~ 37 dB mounted on to the 4 K stage of the cryostat. The signal is further amplified by a room temperature FET amplifier with a gain of ~ 11 dB and finally analysed using either a vector network analyzer or a spectrum analyzer. Sensitivity measurements are carried out by small signal response of the system. RF signals at a frequency fc is applied on to the resonator while biasing the QPC towards an appropriate region of the transconductance trace. A small high-frequency excitation to modulate the channel conductance is applied onto the QPC gate using the home-built bias-tee and the reflected RF signal is acquired using the spectrum analyzer. The sensitivity of the QPC is determined using the formula $\delta G = \frac{1}{2}\frac{\Delta G}{\sqrt{BW}10^{SNR/20}}$, where $\Delta G$ is the conductance change against the small gate excitation signal, $BW$ is the resolution bandwidth used to characterize the signal, and $SNR$ is the signal-to-noise ratio of the sidebands.

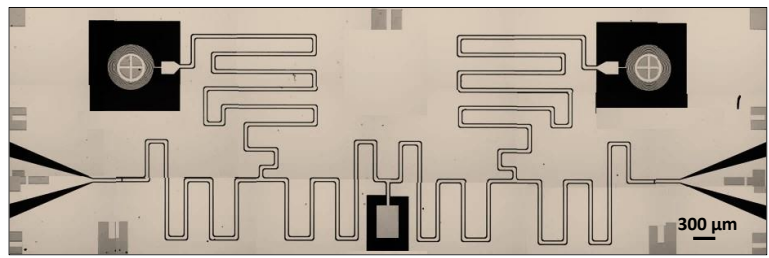

*S-2: Optical image of the resonator similar to the one used in the experiment*

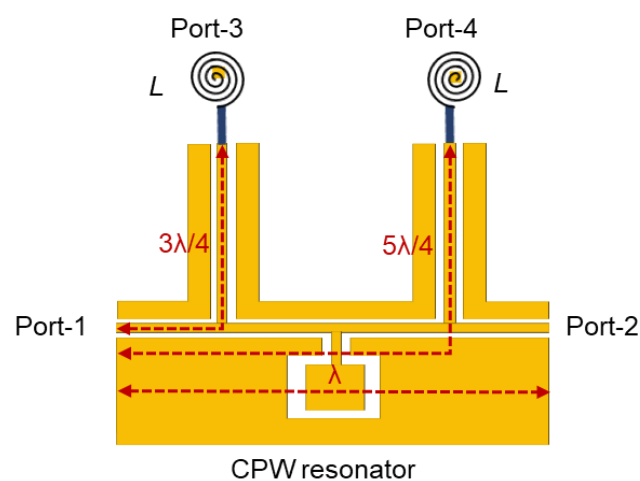

*S-3: Schematic diagram of the resonator showing the length of the segments giving rise to the three modes: the λ, 3λ/4 and 5λ/4 segments, which corresponds to the resonant frequencies 2.930 GHz (1st harmonic), 2.155 GHz (fundamental mode) and 2.285 GHz (1st harmonic), respectively[1].*

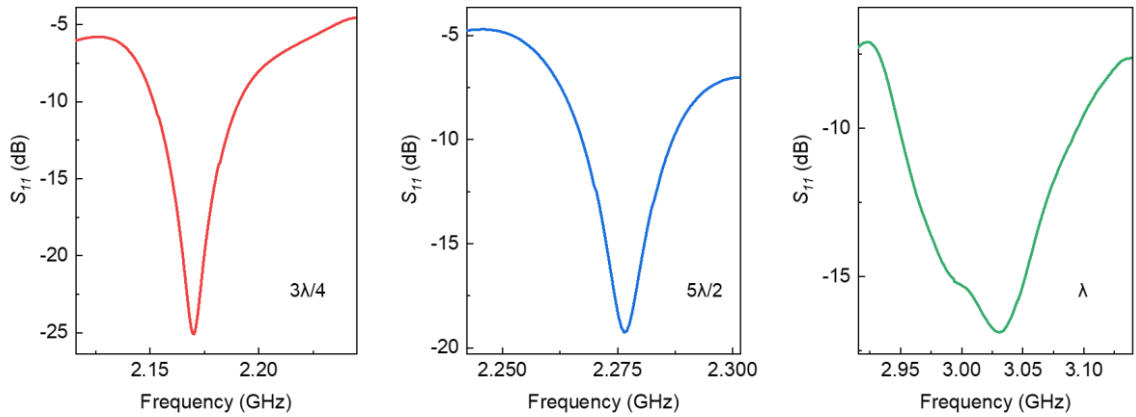

*S-4: $S_{11}$ characteristics of the resonator at three resonances.*

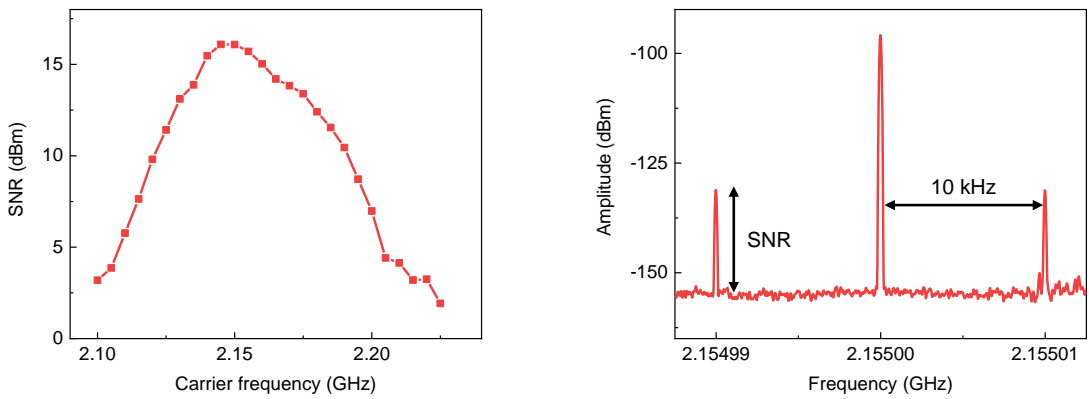

*S-5: Signal to noise ratio (SNR) vs carrier frequency around the resonant frequency, 2.17 GHz in response to a 2 mV$_{RMS}$, 10 kHz gate excitation signal of $P_{rf} \sim -89.19$ dBm. From the plot, it is clear that the signal is maximum for a frequency of 2.155 GHz, which has been used for all the following measurements in this paper, unless stated otherwise. Right hand side plot shows a representative spectrum of the small signal response of the system for an applied gate excitation of 10 kHz.*

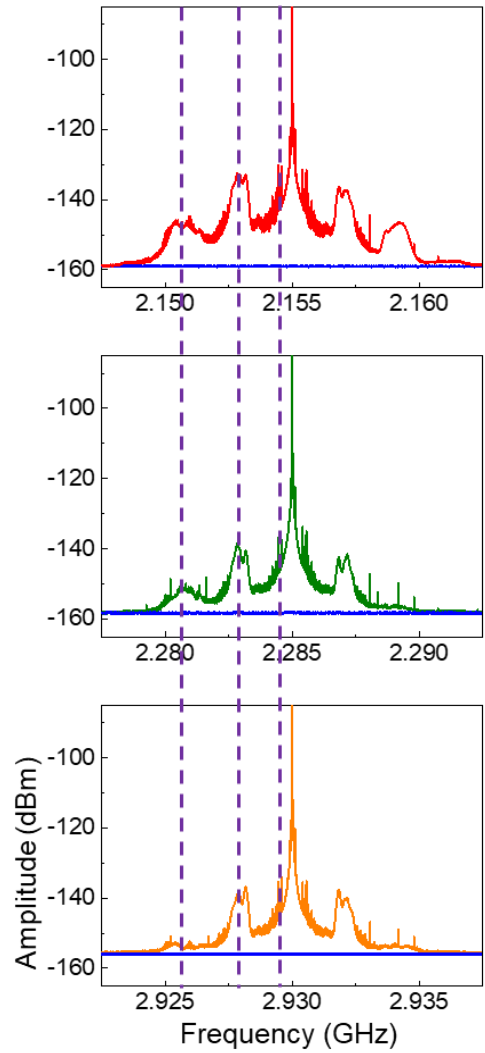

*S-6: Reflected power spectrum at three different carrier frequency 2.155 GHz, 2.285 GHz and 2.930 GHz, respectively. The frequency-dependent noise characteristics are the same in all three cases, i.e., independent of the driving frequency.*

### S-7: Converting Applied power to voltage across QPC

The power reaching the contact pad of the resonator is calculated by subtracting the total attenuation (-49.19 dB) in the input line of cryostat and is converted to Watts.

- The Power stored in the resonator is calculated using the equation[2],

$$P_{stored}(W) = \frac{4Qr(1-r)P_{in}}{\pi}$$

where $Q$ is the quality factor, $r$ is reflection coefficient calculated from $S_{11}$ and $P_{in}$ is the input power in Watts at the contact pad of the resonator.

- Power to voltage conversion, $V = \sqrt{P_{stored} Z_0} = \sqrt{\frac{4Qr(1-r)P_{in}}{\pi} Z_0}$

- Voltage coupled to QPC, (from coupling factor), $V_{QPC} = \lambda_g V = \lambda_g \sqrt{\frac{4Qr(1-r)P_{in}}{\pi} Z_0}$

where, $\lambda_g = \sqrt{Z_0/R} = \sqrt{50/2660} = 0.137$ is the coupling factor.

## S-8: Vibrational modes of the crystal

| Frequency | Vibrational mode type | COMSOL simulation snapshot |
|---|---|---|
| 530 kHz | Flexural mode (x-direction) | |
| 2.078 MHz | Flexural mode (x-direction) | |
| 4.345 MHz | Flexural mode (x-direction) | |

## S-9: Displacement calculation

Displacement $\Delta z$ is calculated using the equation,

$$\Delta z = \frac{e}{\lambda} \sqrt{\frac{\pi \nu_m}{2Q_m} \frac{P_{n,max}}{G_{QPC} \, K \, RBW}}$$

- $e = 1.6 \times 10^{-19} \, C$
- $\lambda = 1.46 \, fN = 1.46 \times 10^{-15} \, N$ (sample-B)
- $\nu_m$ (mechanical resonant frequency), $Q_m$ (Quality factor for the mechanical mode) and $P_{n,max}$ (maximum peak height of the noise peak) are found from spectral decomposition of the noise peaks.

- $G_{QPC}$ is the operating point conductance, calculated from $R_{QPC} = 2.67\ k\Omega$.
- $K$ is the power coupling factor between the resonator-QPC and measurement circuit.

$$K = \lambda_g^2 = \frac{Z_0}{R} = \frac{50\ \Omega}{2.67\ k\Omega} = 0.0187$$

- $RBW$ is the resolution bandwidth, $RBW = 100\ Hz$.

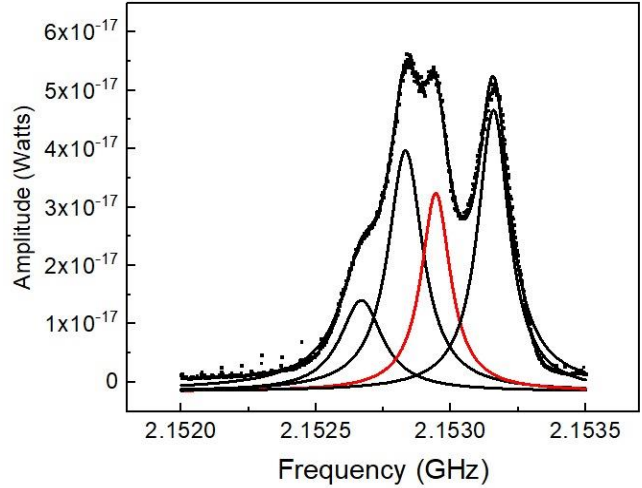

*S-10: Spectral decomposition of the noise peak at 2.15 MHz from the center frequency 2.155 GHz, for the calculation of mechanical Q-factor. The peak marked in red colour has been used for the displacement calculations.*